\documentclass[11pt,a4paper]{article}
\pdfoutput=1

\usepackage{jcappub}
\usepackage{multirow}
\usepackage{amssymb} 
\usepackage[normalem]{ulem}
\usepackage{subfigure}

\newcommand{\dd}{{\rm{d}}}

\newcommand{\be}{\begin{equation}}
\newcommand{\ee}{\end{equation}}

\newcommand{\kop}%{{\cal k}}
{\mathfrak{K}}

\definecolor{darkgreen}{rgb}{0,0.6,0}
\definecolor{darkcyan}{rgb}{0.0, 0.55, 0.55}

\newcommand{\Hc}{\mathcal{H}}

\newcommand{\inspire}[1]{\href{http://inspirehep.net/search?p=find+J+#1}
	{{\color{black}[{\color{blue} {\small in}SPIRE}]}}}
\newcommand{\book}[1]{\href{http://inspirehep.net/search?p=#1}
	{{\color{black}[{\color{blue} {\small in}SPIRE}]}}}

\newcommand{\inspired}[1]{\href{http://inspirehep.net/search?p=#1}
	{{\color{black}[{\color{blue} {\small in}SPIRE}]}}}

\newcommand{\nab}{\nabla}

\makeatletter
\newsavebox\myboxA
\newsavebox\myboxB
\newlength\mylenA

\newcommand*\mybar[2][0.75]{%
	\sbox{\myboxA}{$\m@th#2$}%
	\setbox\myboxB\null% Phantom box
	\ht\myboxB=\ht\myboxA%
	\dp\myboxB=\dp\myboxA%
	\wd\myboxB=#1\wd\myboxA% Scale phantom
	\sbox\myboxB{$\m@th\overline{\copy\myboxB}$}%  Overlined phantom
	\setlength\mylenA{\the\wd\myboxA}%   calc width diff
	\addtolength\mylenA{-\the\wd\myboxB}%
	\ifdim\wd\myboxB<\wd\myboxA%
	\rlap{\hskip 0.5\mylenA\usebox\myboxB}{\usebox\myboxA}%
	\else
	\hskip -0.5\mylenA\rlap{\usebox\myboxA}{\hskip 0.5\mylenA\usebox\myboxB}%
	\fi}
\makeatother

\begin{document}
	
	\title{Relativistic bias in neutrino cosmologies}
	
	\date{\today}
	
	\author[a]{Christian Fidler,}
	\emailAdd{fidlerc@physik.rwth-aachen.de}
	
	\author[a]{Nils Sujata and}
	\emailAdd{nils.sujata@rwth-aachen.de}
	
	\author[a]{Maria Archidiacono}
	\emailAdd{archidiacono@physik.rwth-aachen.de}

	\affiliation[a]{Institute for Theoretical Particle Physics and Cosmology (TTK), RWTH Aachen University, Otto-Blumenthal-Strasse, D--52057 Aachen, Germany}

	\abstract{
		Halos and galaxies are tracers of the underlying dark matter structures. While their bias is well understood in the case of a simple Universe composed dominantly of dark matter, the relation becomes more complex in the presence of massive neutrinos. Indeed massive neutrinos introduce rich dynamics in the process of structure formation leading to scale-dependent bias. We study this process from the perspective of general relativity employing a simple spherical collapse model. We find a characteristic signature at the neutrino free-streaming scale in addition to a large-scale feature from general relativity. The scale-dependent halo bias opposes the suppression in the matter distribution due to neutrino free-streaming and leads to corrections of a few percent in the halo power spectrum. It is not only sensitive to the sum of the neutrino-masses, but respond to the individual masses. Accurate models for the neutrino bias are a crucial ingredient for the future data analysis and play an important role in constraining the neutrino masses.   }
	
	\maketitle   
	
	\flushbottom
	%%%%%%%%%%%%%%%%%%%%%%%%%%%%%%%%%%%%%%%%%%%%%%%%%%%%%%%%%%
	\section{Introduction}
	\label{sec:intro}
	
	Since the first measurement of neutrino oscillations \cite{Fukuda:1998mi}, we know that neutrinos must have a small but non-zero mass. Oscillation experiments have determined the mass splittings with remarkable precision \cite{Esteban:2018azc}, while the absolute masses remain largely unknown. Their small values make it extremely challenging to measure them in direct experiments and cosmology is one of the most promising approaches to constrain or detect the mass of the neutrinos \cite{Aghanim:2018eyx,Archidiacono:2016lnv,Benson:2011uta,Seljak:2006bg}.
	
	Even a small mass of neutrinos modifies the evolution of matter in the Universe and when integrated up over the age of the Universe these tiny changes can lead to observable differences in the distribution of galaxies. In this work we focus on a particular effect that neutrinos have in the process of structure formation; changing the halo bias. Halos and galaxies form from the most overdense collapsing regions and the presence of neutrinos may modify the dynamics of this collapse and lead to a altered abundance of galaxies or other tracers of cold dark matter \cite{Lesgourgues:2006nd,Saito:2009ah,Castorina:2013wga,Castorina:2015bma,Massara:2014kba}.  
	Initially neutrinos are hot and do not support the gravitational collapse of matter, but as the Universe expands they cool down and eventually can be captured by the gravitational potentials of the cold dark matter structures. The epoch when this transition occurs is fixed by the neutrino masses and, given the current bounds, it coincides with the epoch of matter-domination when the observable structures form. 
	From this point onwards neutrinos start to contribute to the the small scale collapse and affect the bias of halos or galaxies with respect to the underlying matter distribution. 
	
	In this paper we describe this process in general relativity using approximations that capture the essential physics, but at the same time are simple enough so that we may employ a semi-analytical approach. Our work is complementary to N-body simulations; it allows for a more detailed physical understanding of the relevant processes while it relies on a series of simplifications that are not needed in a particle simulation. However the inclusion of neutrinos in N-body simulations is still a relatively novel field that comes with its own uncertainties. Our work relies on the following main assumptions and simplifications:
	\begin{itemize}
		\item We employ the weak-field limit of general relativity \cite{Fidler:2017pnb,Brustein:2011dy}.
		\item The small scale collapse is assumed to be perfectly spherical.
		\item We neglect any feedback of the small scale dynamics on the much larger scales.
		\item We assume that the cosmological scales on which we compute the bias are significantly larger than the scales of an individual collapse.
	\end{itemize} 
	The weak field limit of gravity is valid throughout most of the collapse. On the other hand, assuming a spherical over-density is a strong simplification since the initial overdensities are irregular, potato-shaped patches. As soon as non-linearities become relevant in the collapse these are subject to a more complex collapse dynamics, especially including anisotropic stresses. We assume that the impact of neutrinos on the collapse does not differ significantly depending on the shape and employ a perfect spherical symmetry that allows a much simpler treatment of the collapse. We will improve this approximation by employing a halo mass function that was fitted to a N-body simulation not relying on this approximation. 
	The final two assumptions justify the commonly applied picture of the peak-background split \cite{Sheth:1999mn}.
	
	The neutrino bias has been studied using both numerical and analytic methods. N-body simulations can be modified to include neutrinos and study their dynamics in full non-linearity \cite{Castorina:2013wga,Chiang:2017vuk}. However, the simulation of neutrinos in general relativity is still a novel field and often relativistic corrections are neglected, see \cite{Adamek:2017uiq,Dakin:2017idt,Fidler:2018bkg} for recent developments. In \cite{LoVerde:2014pxa,Munoz:2018ajr} a semi-analytic approach very close to our ansatz has been studied, however not considering all relativistic corrections. We aim to derive a treatment that includes all terms in general relativity required to describe light species such as neutrinos accurately. 
	
	The paper is organised as follows. After a brief review of bias, we review the spherical collapse in the framework of general relativity and compare our results with the commonly employed closed-Universe ansatz for collapsing shells in section \ref{sec:collapse}. Then, in section \ref{sec:halo_mass} we discuss the halo-mass function and how it can be fitted to simulations when taking a relativistic point of view. We describe our numerical method in section \ref{sec:numerics} and present our results for selected neutrino masses. Finally we compare our analysis with the literature and conclude.

	%%%%%%%%%%%%%%%%%%%%%%
	\subsection{Definitions}
	We employ the Poisson gauge, with the metric line element
	\be \dd s^2 = g_{\mu\nu} \,\dd x^\mu \dd x^\nu = g_{00}\, \dd \tau^2 + 2 g_{0i}\, \dd x^i \dd \tau + g_{ij} \,\dd x^i \dd x^j \,,
	\ee
	given by the scalar type metric perturbations
	\begin{subequations}
		\label{metric-potentials}
		\begin{align}
		g_{00} &= -a^2 \left[ 1 + 2 \Psi \right] \,, \\
		g_{0i} &=  0 \,,\\
		g_{ij} &= a^2 \left[ \delta_{ij} \left( 1 + 2 \Phi \right)\right] \,.
		\end{align}
	\end{subequations}
	Here, $a$ is the cosmic scale factor, $\delta_{ij}$ the Kronecker symbol and we make use of the conformal time $\tau$.
	The Poisson gauge has several advantages for our analysis (see also the discussion in \cite{Fidler:2017pnb}). It is compatible with the assumptions of weak-field relativity up to the small scales. It furthermore does not introduce shape-changing distortions in the spatial sector. This implies that an initially spherical setup will remain spherically symmetric when expressed in Poisson gauge coordinates.

	\section{A brief review of bias in general relativity}
	\label{sec:bias_review}
	The bias is defined as the relation between structures and the underlying matter density \cite{Sheth:1999mn}. We write a linear Eulerian bias in Fourier space as 
	\be
	\delta_h(k) = b(k) \delta_{cdm}(k)\,,
	\ee
	where $\delta_h = \frac{n - \bar{n}}{\bar{n}}$ is the over-density of a population of tracers and $\delta_{cdm}$ the fractional cold dark matter overdensity.   
	The bias $b$ is the result of a complex dynamic leading to the collapse, including contributions from all scales. These include the small-scale dynamics of the gravitational collapse and the impact of the much larger modes on the small scale evolution. Furthermore the Eulerian bias receives a contribution from the evolution of the matter perturbations as the tracers are comoving with the dark matter.
	
	To separate these different contributions we analyse the collapse in Lagrangian space, where our observer is chosen to be comoving with a collapsing cloud of matter. In these coordinates we will compute the Lagrangian bias
	\be
	\delta_h = b_{\rm Lag} \delta_{cdm}\,,
	\ee
	where $\delta_{cdm}$ refers to the initial matter overdensities and $\delta_h$ to the halo density in Lagrangian coordinates. This Lagrangian bias $b_{\rm Lag}$ isolates the bias in the formation of the tracer and is connected to the Eulerian bias via
	\be
	b_{\rm Lag} = b + 1\,.
	\ee 
	This relation expresses nothing but the simple statement that any object that formed unbiased with respect to dark matter ($b_{\rm Lag} = 0$) will still end up being biased in Eulerian space because it follows the dark matter evolution and therefore is a perfect tracer of dark matter ($b = 1$). 
	
	The Lagrangian bias can be expressed as
	\be
	b_{\rm Lag}(z) = \frac{d \log n(z)}{d \delta_{cdm}}\,,
	\ee
	stating that the bias is simply the change in the number density of the structures that have formed up to redshift $z$ compared to the underlying dark matter long-mode overdensity. 
	
	We introduce the critical density $\delta_{\rm crit}(z)$ as the small-scale initial density required to form a halo by redshift $z$. This density in general does depend on the large modes $\delta_{cdm}$ that influences the dynamics of the collapse. We then find:
	\be \label{eq:split}
	b_{\rm Lag}(z) = \frac{d \log n(z)} {d \delta_{\rm crit}(z)} \frac {d \delta_{\rm crit} (\delta_{cdm})}{d \delta_{cdm}}\,,
	\ee
	where all perturbations should be understood as evaluated at the initial time. In particular $\delta_{\rm crit}(z)$ is not a critical density at redshift of $z$, but the initial overdensity required for a collapse until the redshift of $z$. The long mode $\delta_{cdm}$ represents the matter content of the Universe at the initial time, but since we assume adiabatic initial conditions, setting a particular value of $\delta_{cdm}$ automatically fixes the long mode perturbations of all species. In that sense this derivative can be understood as the change of the critical density with respect to a change in the initial long mode curvature perturbation. 
	
	Equation \ref{eq:split} is crucial for our calculation since it separates the physics into a large and a small scale part. The first term is known as the halo mass function, describing the amount halos that may be formed from the initial small scale perturbations under the assumptions that a certain minimal density $\delta_{\rm crit}(z)$ is required to form a halo. It is a statistical quantity determined by the primordial power spectrum and the small scale physics. In contrast, the second term describes how a large mode affects the collapse by changing this critical threshold to form a halo. This separation is known as the peak-backgorund split. 
	We will now study both contributions to the bias separately in general relativity.
	
	\subsection{Spherical collapse}\label{sec:collapse}
	The relation between the initial small scale density that is required for a collapse and the long wave-length modes can be approximated by looking at the simplified problem of a spherical small-scale overdensity embedded in a long-wavelength perturbation. Since in linear theory modes evolve independently it is sufficient to look at each long mode in isolation. We will further assume that the Fourier wavevector $k$ of the large mode is much larger than the the scale of collapse: $k \ll k_{\rm collapse}$. In this case the long mode may be considered spatially flat and acts similar to a modification of the background, only that it features a different time evolution and scale-dependencies.   
	
	First we study the evolution of a small scale mass-shell in full non-linearity including general relativity in the weak field limit. We do not assume that the small scale densities or velocities are small, only the metric potentials are assumed to be small and higher orders in them may be neglected.
	We separate the metric potential into the contribution from the large mode and the spherical shell: $\Phi = \Phi^{\rm S} + \Phi^{\rm L}$, where $S$ labels the small-scale shell and $L$ is the long wavelength mode. Note that the long mode may be locally constant, but being a perturbation it has a different time evolution than a background quantity. We employ linear perturbation theory to compute the long mode perturbations according to adiabatic initial conditions. 
	
	The small scale metric potentials in weak field gravity may be computed directly from the shell's overdensity
	\be
	-\nabla \Phi^S = 4 \pi G a^2 \delta\rho^S\,,
	\ee
	and $\Phi^S = -\Psi^S$ since a spherical configuration does not posses anisotropic stresses.  
	
	We define the total mass of our shell
	\be
	M = \int \limits_0^{r_{\rm S}} d^3r a^3 (1 + 2 \Phi)^{\frac 3 2} \rho\,,
	\ee 
	where $r_{\rm S}$ is the radius of the shell and $\rho = \bar{\rho} (1 + \delta_S +\delta_L)$ is the total density including the long mode and the background density. The mass is a very useful quantity as it is preserved in the collapse. The gravitational potential is not sensitive to the total mass but is sourced from the overdensities alone. We define the mass difference with respect to the average mass in a shell of the given radius from background and long mode contributions: 
	\be
	\Delta M = \int \limits_0^{r_{\rm S}} d^3r a^3 (1 + 2 \Phi)^{\frac 3 2} \delta\rho_S\,,
	\ee
	that is only sourced by the small-scale shell overdensity $\delta\rho_S$. Using this definition we can express the potential as
	\be
	\Phi^{\rm S}(r) = \frac {\Delta M G}{a r}\,,
	\ee
	providing the well known Newtonian result that we find here since weak-field relativity has a Newtonian limit on the small scales. 
	
	Due to the spherical symmetry of our problem the equations of motion may be derived from the geodesic equation of a particle sitting at the edge of the shell, identifying its position with the radius of the shell:
	\be \label{GR:shell}
	\ddot{r} + 2 (\Hc + \dot{\Phi})\dot{r} + \frac {\Delta M G}{a r^2} = 0\,,
	\ee
	where we have removed all contributions that simply displace the shell, i.e. they act on a particle at position $+r$ in the same way as at $-r$. These terms are needed in an Eulerian description, but can be dropped since our observer is assumed to be comoving with the shell.
	The term $\frac {\Delta M G}{a r^2}$ is driving the collapse via the shell's own gravitational potential,
	while $2 (\Hc + \dot{\Phi})\dot{r}$ describes the Hubble drag that is working against the collapse by reducing the particles' velocities. Note that the Hubble drag receives a relativistic correction related to the local expansion of space caused by the shells own gravitational potential and the long mode evolution from $\dot{\Phi}$.
	
	In the literature the closed-Universe ansatz is often employed to describe the collapse dynamics instead of solving the relativistic geodesic equation \ref{GR:shell}. In this case the over-dense shell is described as its own miniature Universe and since it is completely isotropic the collapse can be described by the Friedmann equation of that local curved Universe. The evolution of the scale factor $\frac{\ddot{a}}{a}$ is then identified with the ratio of the shell's radius $\frac{\ddot{R}}{R}$.  
	This leads to the alternative collapse equation:
	\be
	\frac{\ddot{R}}{R} + \frac{M G}{R^3} = 0\,,
	\ee
	where the mass is defined by $M = \frac{4\pi}{3}\rho R^3$. In order to compare the two equations one first needs to understand the differences in the coordinates. In our relativistic equation $r$ is a coordinate distance while $R$ in the closed Universe ansatz is a physical radius. Furthermore the time coordinate of our equation is the conformal time defined with respect to the background Universe $\tau$, while the time coordinate of the closed Universe case is the proper time of the shell alone. 
	
	We recast our equation into physical radius and time to find:
	\be\label{eq:phy_coord}
	\frac{\ddot{R}}{R} + \left( \frac{\dot{a}}{a} + \dot{\Psi} \right)\left( \frac{\dot{R}}{R} - \frac{\dot{a}}{a} \right) - \ddot{\Phi} + 3 \frac{\dot{a}}{a} \dot{\Phi} + \frac {\Delta M G}{ R^3} - \frac{\ddot{a}}{a}  = 0\,,
	\ee
	where for this equation ovetdots refer specifically to the time coordinate of the Poisson gauge and not the conformal time $\tau$ as otherwise assumed throughout the paper. 
	
	The most striking differences between the closed Universe case and our analysis is that the collapse seems to be driven by the total mass and not only the local mass difference. However, this is quickly reconciled using the Friedmann equation, at least in a pure matter Universe where $\frac {\Delta M G}{ R^3} - \frac{\ddot{a}}{a}$ in Eq.~(\ref{eq:phy_coord}) is exactly the total mass.
	
	This leaves a few relativistic corrections and a term describing a Hubble drag: 
\be
	 \left( \frac{\dot{a}}{a} + \dot{\Psi} \right)\left( \frac{\dot{R}}{R} - \frac{\dot{a}}{a} \right)\,.
\ee
	 This term is working against the collapse and starts to be relevant when the shell deviates from the background evolution. As a consequence this term is initially vanishing but it becomes more relevant towards the later stages of the collapse. 
	In the closed Universe case such a term cannot appear as the shell is treated as its own Universe, i.e. it is always comoving with its local Hubble rate. 
	
	The two descriptions may seem contradictory at first glance, but are brought into agreement when realising the difference in their respective time coordinates. As long as the shell is comoving with the Hubble flow of the background Universe, the local time coordinate matches the Poisson gauge time. But as long as this is the case the Hubble drag term discussed above also vanishes and both descriptions are identical.
	At later stages when the shell starts to decouple from the Hubble flow, both approaches also employ different time coordinates. While we employ the usual Poisson gauge time, the closed Universe case uses the proper time of the shell. At the moment of collapse $(R=0)$, the time of the local patch even becomes singular while the time of the background Universe is not affected by this event. 
	The additional relativistic corrections $- \ddot{\Phi} + 3 \frac{\dot{a}}{a} \dot{\Phi}$, representing time dilation effects, are explained in the same way and vanish in terms of the proper time of the shell. 
	
	\begin{figure}[htbp]
		\centering
		\includegraphics[width=0.95\textwidth]{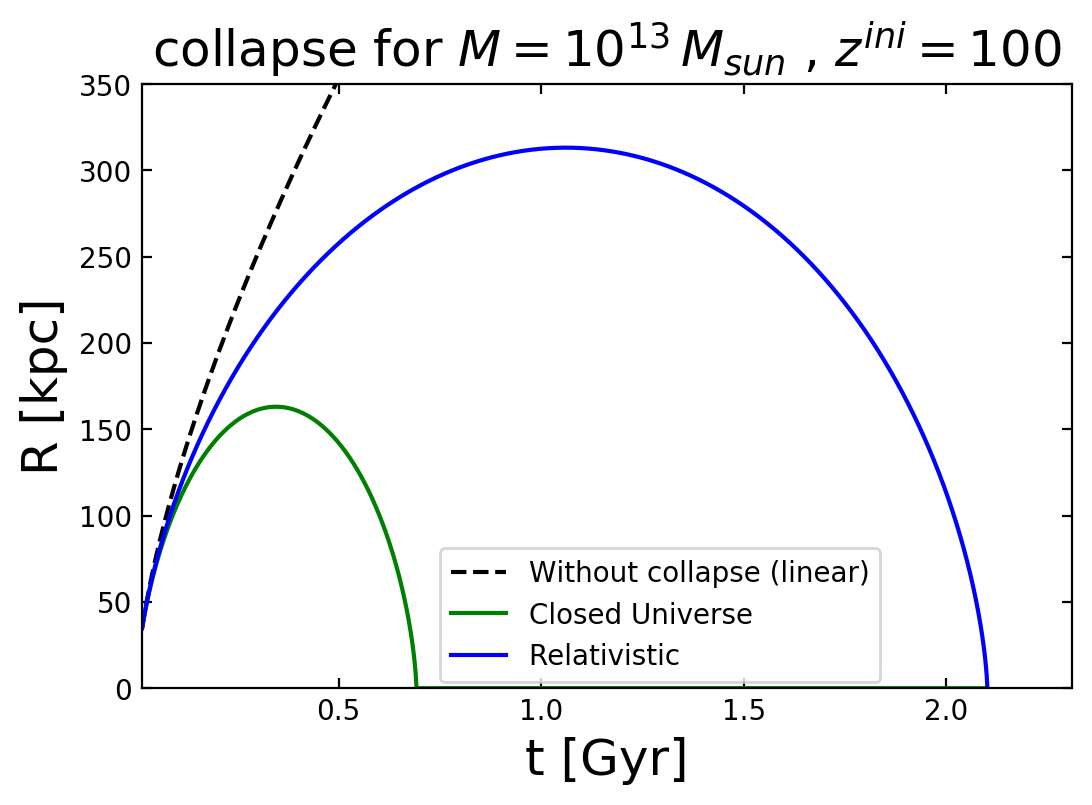}
		\caption{The different collapse histories using the geodesic equation versus the closed Universe case and linear theory. Initially all three description agree but as the collapse starts to decouple form the background flow the closed Universe case deviates due to the different time coordinate. Around the same time non-linear corrections become relevant causing the departure from the linear solution.}
		\label{fig:collapse_comp}
	\end{figure}
	
	The different collapse timelines are shown in Figure \ref{fig:collapse_comp} with respect to their own time coordinates. Even though the Hubble drag is a small correction at any given time, it accumulates over the collapse history and significantly delays the collapse. The relativistic corrections are much smaller but still non-negligible. 
	
	In a pure matter case both descriptions are equivalent, but one needs to remember that the closed Universe case employs the time observed in the local collapsing patch. This is particularly important when comparing the results to a linear Boltzmann code that employs Poisson or synchronous gauge time coordinates.
	
	In the case of a Universe that has a small radiation component or a cosmological constant, the closed Universe ansatz fails. The reason is that the time evolution of all perturbations in the closed Universe case are automatically assumed to be identical and scale-independent. This is particularly problematic when embedding the spherical collapse in a long mode perturbation. As a consequence we will find that the long modes of the different species enter in our equations in a different way than in the closed Universe case, and only for a matter Universe these agree. In addition the closed Universe ansatz assumes that all particles initially inside the patch will remain there indefinitely and follow the collapse, which is not the case for relativistic species such as neutrinos or photons.
Indeed, if the closed Universe calculation is employed in a non-pure matter Universe, it leads to the collapse equation \cite{LoVerde:2014pxa}
		\be\label{eq:closed-rad}
		\frac{\ddot{R}}{R} + \frac{M G}{R^3}  + \frac{4\pi G}{3} [\rho_\gamma + 3 p_\gamma - 2 \rho_\Lambda]= 0\,,
		\ee
		where all sources of energy contribute to the collapse in the same way.
		On the contrary in our geodesic Equation \ref{GR:shell} radiation perturbations do not contribute directly to the collapse and only indirectly act via the relativistic corrections.
	
	In \cite{Hu:2016ssz} it has been argued that the problem of scale-dependent evolution of long modes can be avoided by matching the physical problem to a specifically designed spherical collapse instead of using Eq.~(\ref{eq:closed-rad}). While the long modes can then include various species it is still assumed that the small mode within the closed Universe is composed of only collapsing cold dark matter. This method has been applied in \cite{Munoz:2018ajr} where the impact of the neutrinos on the long mode is consistently included in the spherical collapse. Even though neutrino perturbations never become large and contribute to clustering only at the background level \cite{Castorina:2013wga}, the feedback of the cold dark matter non-linear evolution on neutrinos may be non-negligible. We do account for this effect by not employing the closed Universe ansatz and instead solving directly the relativistic geodesic equation.
	
	In addition to the above discussion one needs to be very careful when using Eq.~(\ref{eq:closed-rad}). In the closed Universe case there is no notion of a gauge, even though the long mode perturbations are explicitly gauge dependent. In the pure matter case this problem was avoided by using only the physical radius $R$ and the total mass $M$, which are both gauge independent quantities. However, when including other perturbations one needs to argue in which gauge these should be understood.
	
	Not taking these issues into account may lead to results not consistent with general relativity. For example due to the mismatch of the time coordinates used to describe the collapse and the time coordinate employed in a Boltzmann code, neutrinos would be evaluated at far too early times when they may still be relativistic and their impact on the collapse would be underestimated. We believe that these complications are responsible for the differences in the results we find compared to \cite{LoVerde:2014pxa,Munoz:2018ajr}.
	
	\subsection{The halo mass function}\label{sec:halo_mass}
	
	The halo mass function describes how many halos of a given mass will form from the initial small scale perturbations under the assumptions that a certain initial overdensity $\delta_{\rm crit}$ is required for a collapse until redshift $z$. While this is a very simplified picture, it allows for an analytical solution to the halo mass function, the Press-Schechter formalism \cite{Press:1973iz}. We assume that the initial overdensity is described by random primordial fluctuations, with the probability $p_{\delta \geq \delta_{crit}}$ to find a local density above the critical density:
	\be
	p_{\delta \geq \delta_{crit}} = \int_{\delta_{crit}}^{\infty} d \delta \, p_{\delta} = \frac{1}{\sqrt{2 \pi \sigma^2}} \int_{\delta_{crit}}^{\infty} d \delta \, \text{exp}\left(\frac{\delta^2}{2 \sigma^2}\right) = \frac{1}{2} \text{erfc}\left(\frac{\delta_{crit}}{\sigma}\right)\,,
	\ee
	with $\sigma^2(z, M) = \int \frac{dk}{2 \pi^2} \, k^2 \, W(k \, r) \, P_{cc}(k, z)$ the variance for fluctuations of total mass $M$. Here $P_{cc}$ is the cold dark matter auto power spectrum and $W(k \, r) = \frac{3}{(k \, r)^3} \left[\text{sin}(k \, r) - k \, r \, \text{cos}(k \, r)\right]$ is a top hat window function of the given radius. In the above equation all terms are evaluated at the initial time. 
	
	Using this probability we can directly compute the number of halos in a given mass interval
	\be
	\frac{d n}{d M} = \sqrt{\frac{2}{\pi}} \frac{\rho_m}{M} \frac{d \, \text{log}(\sigma^{-1})}{d M} \nu \, \text{exp}(-\frac{\nu^2}{2}) = \sqrt{\frac{2}{\pi}} \frac{\rho_m}{M} \frac{d \, \text{log}(\sigma^{-1})}{d M} \, f(\nu)\,,
	\ee
	where $f(\nu) = f\left(\frac{\delta_{crit}}{\sigma}\right)$ is the halo mass function that describes the overall halo mass distribution. We have further defined the ratio between the critical density and the variance of the primordial fluctuations $\nu = \frac{\delta_{crit}}{\sigma}$. 
	
\begin{figure}[htbp]
		\centering
		\includegraphics[width=0.95\textwidth]{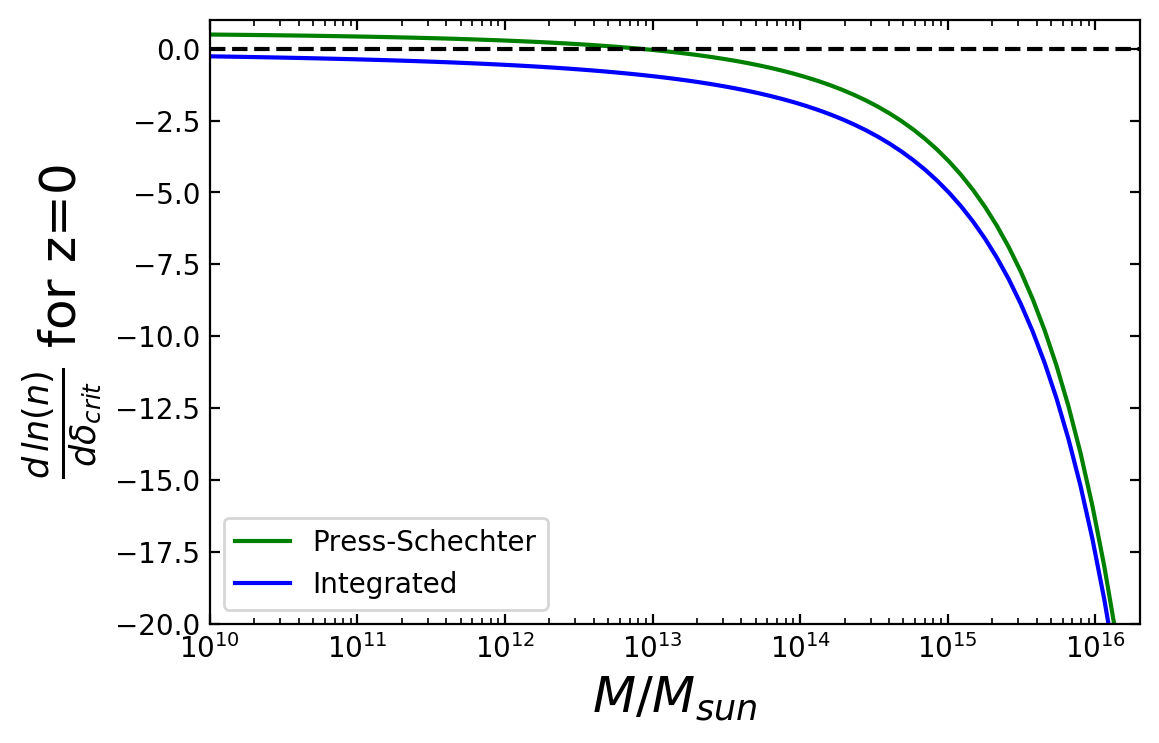}
		\caption{The different mass functions plotted over M. Here "Integrated" refers to Equation (\ref{eq:mass_func_int})}
		\label{fig:massf_comp}
	\end{figure}

	For the computation of the bias we are interested in the change of the halo number with respect to the critical density which is evaluated as
	\be
	\frac{d \, log(n)}{d \delta_{\rm crit}} = - \sqrt{\frac {2}{\pi \sigma^2}} \frac{{\rm exp}(-\frac {\nu^2}{2})}  {{\rm erfc}(\nu) }\,.
	\label{eq:mass_func_int}
	\ee
	This formula is different from the one usually employed in the literature, where the result from \cite{Sheth:1999mn} is taken and adapted to the Press-Schechter case leading to
	\be
	\frac{d \, log(n)}{d\delta_{\rm crit}} = - \frac{1- \nu^2}{\sigma \nu}
	\ee
	This simplification is accurate for high mass halos but eventually overestimates low mass halos, i.e. increasing the critical density would falsely predict that more halos should form. 
	We compare both approaches in Figure \ref{fig:massf_comp}.
	
	In the closed Universe case a very interesting simplification appears. When computing the matter overdensity that a linear code predicts at that time of collapse one finds a rather universal value of $\delta_{\rm linear} = 1.68$, independent of the mass $M$ or redshift of collapse $z$. Often this is interpreted as the value of the overdensity in linear theory at the time when the collapse occurs in the non-linear case. However, in section \ref{sec:collapse} we have discussed that the time arguments of the collapsing regions are significantly offset with respect to the Poisson gauge time used in linear theory. The value of $1.68$ is therefore corresponding to a time significantly before the collapse as seen in Figure \ref{fig:collapse_comp}.
	
	Nevertheless this is a very interesting result as this value does only weakly depend on the initial overdensity leading to the collpase. We further notice that the ratio $\nu = \frac{\delta_{\rm crit}}{\sigma}$ is almost constant in time when using a linear approximation and defining $\sigma$ with respect to the cold dark matter plus baryons only. It may therefore be evaluated at any given time and we can compute it when $\delta_S$, the short wavelength perturbation, is $1.68$ providing $\nu \approx \frac{1.68}{\sigma(\tau_{\rm collapse})}$. 
	In this way we are able to remove the dependance on $\delta_{\rm crit}$, but notice that $\sigma(\tau_{\rm collapse})$ should be evaluated at the time corresponding to the collapse in the closed Universe, which does not correspond to the observed redshift $z$ of the halo. The same would not be possible when using the total matter instead of the cold dark matter only which adds an extra argument why a more universal form is found when using only the cold dark matter as a reference \cite{Castorina:2013wga}.
	
	We define the linear overdensity at the moment of collapse $\delta_{\rm crit}(\tau_{\rm collapse})$ and find from our simulations that
	\begin{eqnarray}
	\delta_{\rm crit}^{\text{closed Universe}}(\tau_{\rm collapse}) &=& \phantom{-}0.21 \delta_{\rm crit} + 1.66 \,, \\
	\delta_{\rm crit}^{GR}(\tau_{\rm collapse}) &=& -4.71 \delta_{\rm crit} + 4.86 \,,
	\end{eqnarray}
	in their respective time arguments. The critical over-density at collapse generally depends on the initial overdensity $\delta_{\rm crit}$ and is not universal. But in the closed Universe case this dependance is very small and leads to the well known value of $1.68$ for an initial $\delta_{\rm crit} \approx 0.1$. Using our relativistic Equation \ref{GR:shell}, $\delta_{\rm crit}(\tau_{\rm collapse})$ is generally much larger and obtains a dependance on the initial overdensity. 
	
	Employing the Press-Schecter formalism with a changed critical density will provide a very different mass function, but we know that it only represents a very simplified model. For this reason the Seth-Tormen formula \cite{Sheth:1999mn}, as well as other more complex fitting functions are usually employed. These model the process of structure formation using N-body simulations thus providing more realistic halo mass functions. The simulation result is then fitted by a simple function motivated by the theoretical arguments above. 
	
	In these fits one parameter is usually defined as a rescaling of $\nu$ and thus any value of $\delta_{\rm crit}$ can be accommodated by simply changing this parameter. An equally good fit can be reached when using $\delta_{\rm crit}(\tau_{\rm collapse}) =1.68$ or any other value. In this way when using a fitting formula the value of $\delta_{\rm crit}(\tau_{\rm collapse})$ is not relevant anymore and any choice will be overwritten by the fit.  
	
\begin{figure}[htbp]
		\centering
		\includegraphics[width=0.95\textwidth]{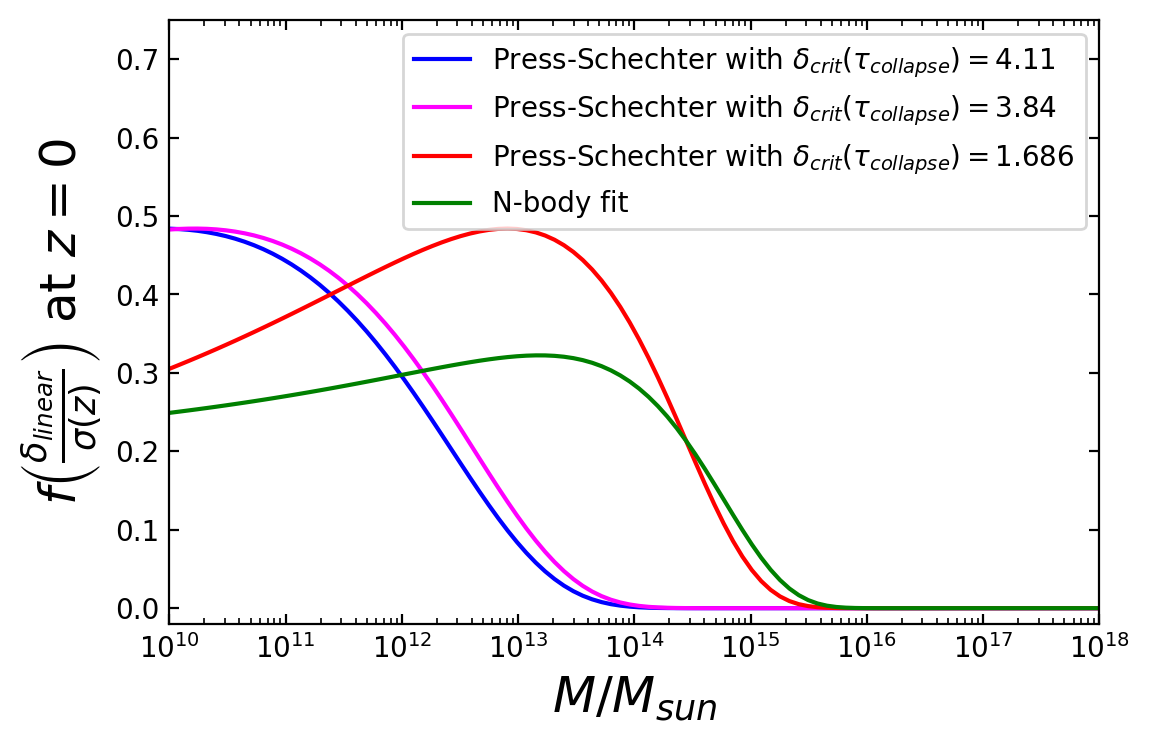}
		\caption{Comparisons of the numerical fit and the Press-Schecter mass functions for various values of $\delta_{\rm linear}$. Simply changing $\delta_{\rm linear}$ does not improve the fit because the model is too simplified to capture the physics. Any value of $\delta_{\rm crit}$ is compatible with the numerical N-body results when a fitting formula is used.}
		\label{fig:mass_function}
	\end{figure}

	We use the mass function proposed in \cite{Bhattacharya:2010wy}, as it is shown in Figure \ref{fig:mass_function} 

	\subsection{A more realistic shell evolution}
	While the above methods already allow us to estimate the bias of structures, we now relax the spherical shell assumption. This will be necessary to accurately simulate neutrinos and to further improve our description of cold dark matter.
	
	So far we have assumed that our initial configuration is a spherically symmetric mass-shell that evolves into increasingly dense shells of constant internal density. However an initially sharply confined mass shell will smooth out during the evolution creating a softer density profile. In particular around the shell a slightly under-dense region will form due to the relativistic evolution. 
	
	We will still assume that our initial configuration is a spherical shell of constant density, but because we are not using a closed Universe ansatz we can allow our shell to evolve into more complex density profiles afterwards. This process is described by the relativistic Navier-Stokes equations for a massive fluid, which can be truncated at the second moment since the spherical symmetry guarantees a vanishing anisotropic stress:
	\be
	(\partial \tau + 3 \Hc) \delta\rho = -3 \rho \dot{\Phi} + \nab_i ( \rho v^i ) \,,
	\ee
	\be
	(\partial \tau + 4 \Hc) (\rho v^i)  - \rho \nab^i \Psi = - 3 \rho v^i \dot{\Phi} + \nab_j (\rho v^i v^j ) \,,
	\ee
	assuming a massive species. For the full equations in the weak-field limit we refer to the computations in \cite{Fidler:2017pnb}. Note that in the case of cold dark matter corrections from non-spherical collapses will have a much bigger impact on the collapse history than this correction. However, the method will be crucial to describe the impact that neutrinos have on the collapse.
	
	\begin{figure}[htbp]
		\centering
		\includegraphics[width=0.95\textwidth]{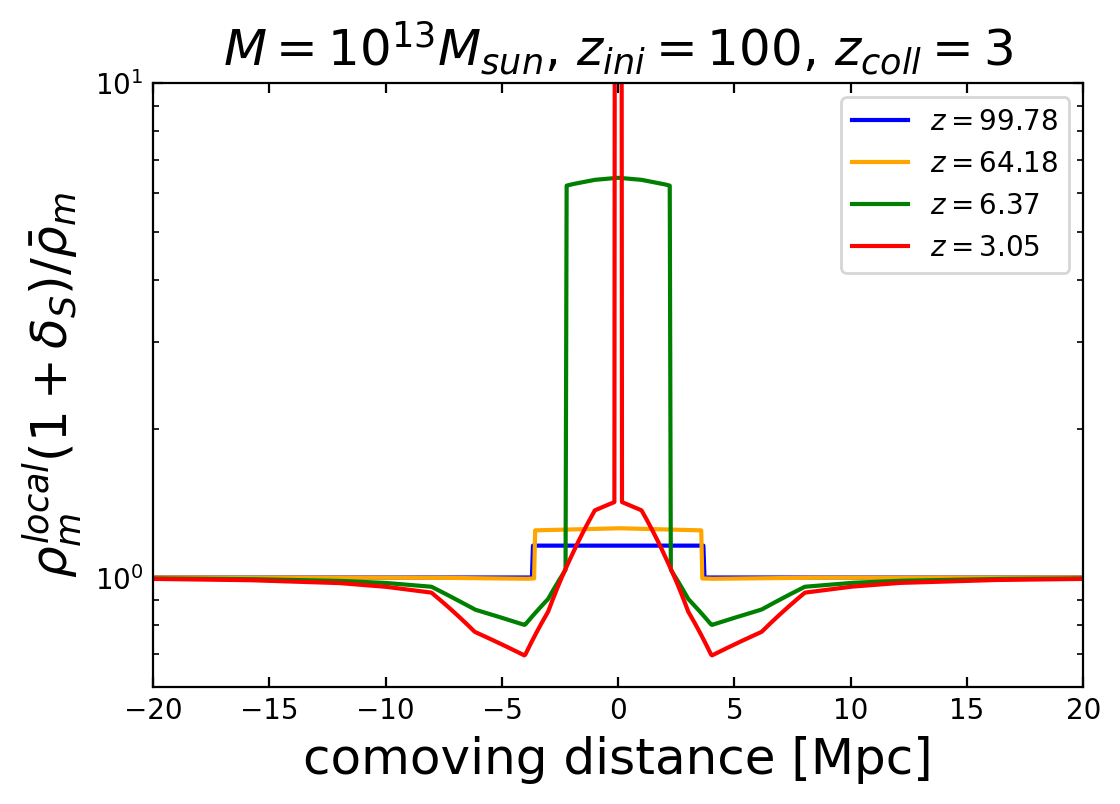}
		\caption{The evolution of an initially spherical mass shell. The density profile slowly develops a gradient towards the centre of the shell. At the same time not enough matter can flow in from the outside causing a small dip in the surrounding matter densities. Note that the last redshift of $3.05$ directly before the collapse shows an artificial density spike that is caused by an insufficient sampling in the radial direction. But since this happens only in the very last stage of the collapse it has a vanishing impact on our results.}	
		\label{fig:mass_shell}
	\end{figure}
	We show the dark matter evolution in Figure \ref{fig:mass_shell}. The initially sharp density profile begins to soften. Around the shell a slight underdense region develops, while on the inside the matter accumulates towards the centre.
	
\begin{figure}[htbp]
		\centering
		\includegraphics[width=0.95\textwidth]{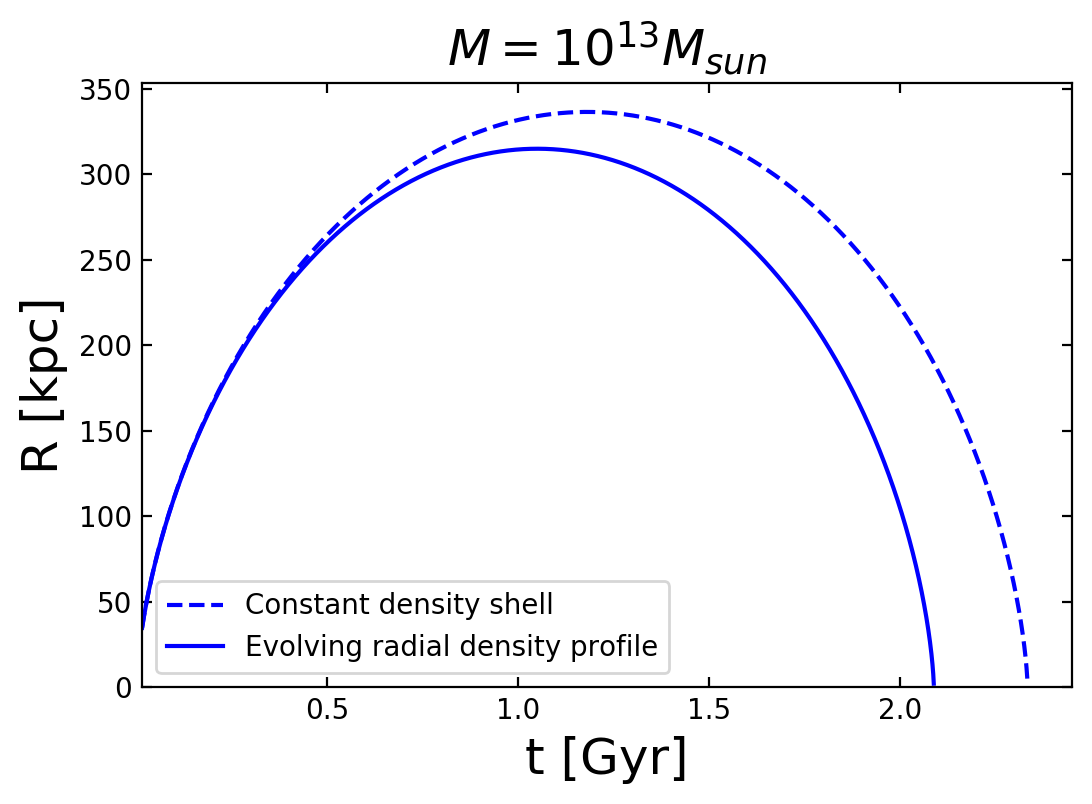}
		\caption{The collapse assuming a constant density shell versus the more realistic assumption of an evolving radial density profile from identical initial conditions. Initially both evolve at the same rate, but the small changes in the density profile eventually leading to an accelerated collapse.}
		\label{fig:coll_background}
	\end{figure}

	This new density profile does affect the collapse dynamics as shown in Figure \ref{fig:coll_background}. We find that the collapse is accelerated due to the increased density gradient resulting from the underdense layer around the shell.
	
	When computing the impact of neutrinos on the collapse this technique is crucial as neutrinos do not follow the massive shell. Instead particles constanly enter and leave the collapsing region while the mean density increases due to the potential of the small-scale dark matter overdensity. Even when the neutrinos are cold, they still have large streaming velocities that work against the collapse. Using our fluid approach we are able to capture the neutrino dynamics and include the impact of their pressure.
	
	\section{Numerical implementation}\label{sec:numerics}
	The computation of the bias is composed of the two separate calculations presented in section \ref{sec:collapse} and \ref{sec:halo_mass}. For the mass function we employ a numerical fit to a N-body simulation, updated to match our relativistic values for $\delta_{\rm crit}$. 
	
	We make use of the CLASS code \cite{Lesgourgues:2011re,Blas:2011rf,Lesgourgues:2011rh} to obtain the linear evolution of the long mode perturbations starting from a shared adiabatic initial mode. We then embed a suite of independent spherical shell simulations in the precomputed long mode while we vary the initial shell density and the overall amplitude of the adiabatic long mode. We evolve these isolated shells to fill a database of independent collapses corresponding to different redshifts, masses and initial conditions. Interpolating these for a fixed mass and redshift we obtain a relation between the initial shell overdensity and the amplitude of the long mode from which the bias can be computed together with the corresponding mass function. 
	
\begin{figure}[htbp]
		\centering
		\includegraphics[width=0.95\textwidth]{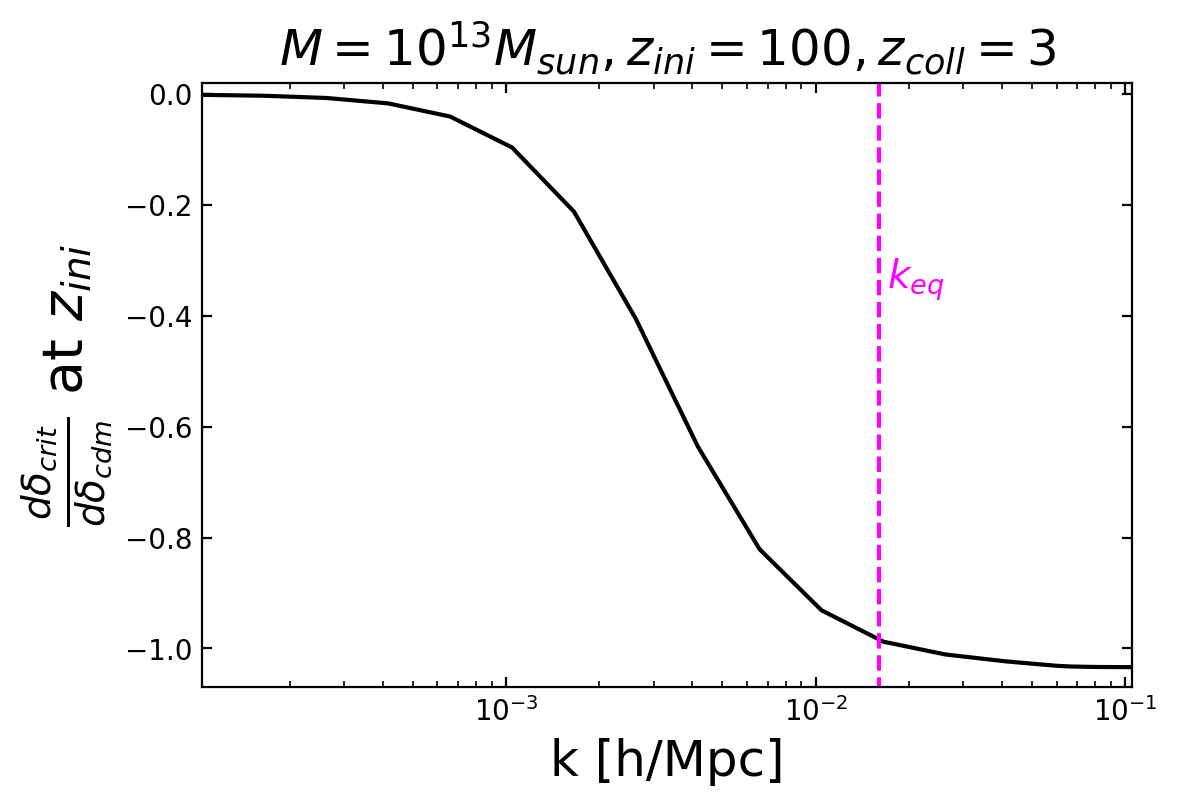}
		\caption{Relation between the long mode and the initial shell density for an object that collapses at redshift $z=3$. A significant rise can be observes on scales larger than $k_{\rm eq}$.}
		\label{fig:matter-crit}
	\end{figure}

	In this work we fix the structure mass to $M = 10^{13} M_{sun}$, the initial time to $z_{ini} = 100$ and the time of collapse to $z_{coll} = 3$. We vary the initial conditions in the range $\delta_{crit} \in [0.14, 0.15]$ (at the initial time) and $\delta_{cdm} \in [0.01, 0.02]$.
	The results are shown in Figure \ref{fig:matter-crit}. On the smaller scales the derivative $\frac{d \delta_{crit}}{d \delta_m} (z_{ini})$ drops and remains close to minus one. This implies that the physics becomes indifferent to the scale of the long mode and the collapse only responds to the total density. This is the expected dynamics in the Newtonian theory. 
	But for much larger modes beyond the matter-radiation-equality scale we find that the relativistic dynamics changes. The long modes become less efficient until they completely decouple from the shell at superhorizon scales. The reason is the different time-evolution of the long modes should they enter the horizon during radiation domination, where they quickly decay. 
	
\begin{figure}[htbp]
		\centering
		\includegraphics[width=0.95\textwidth]{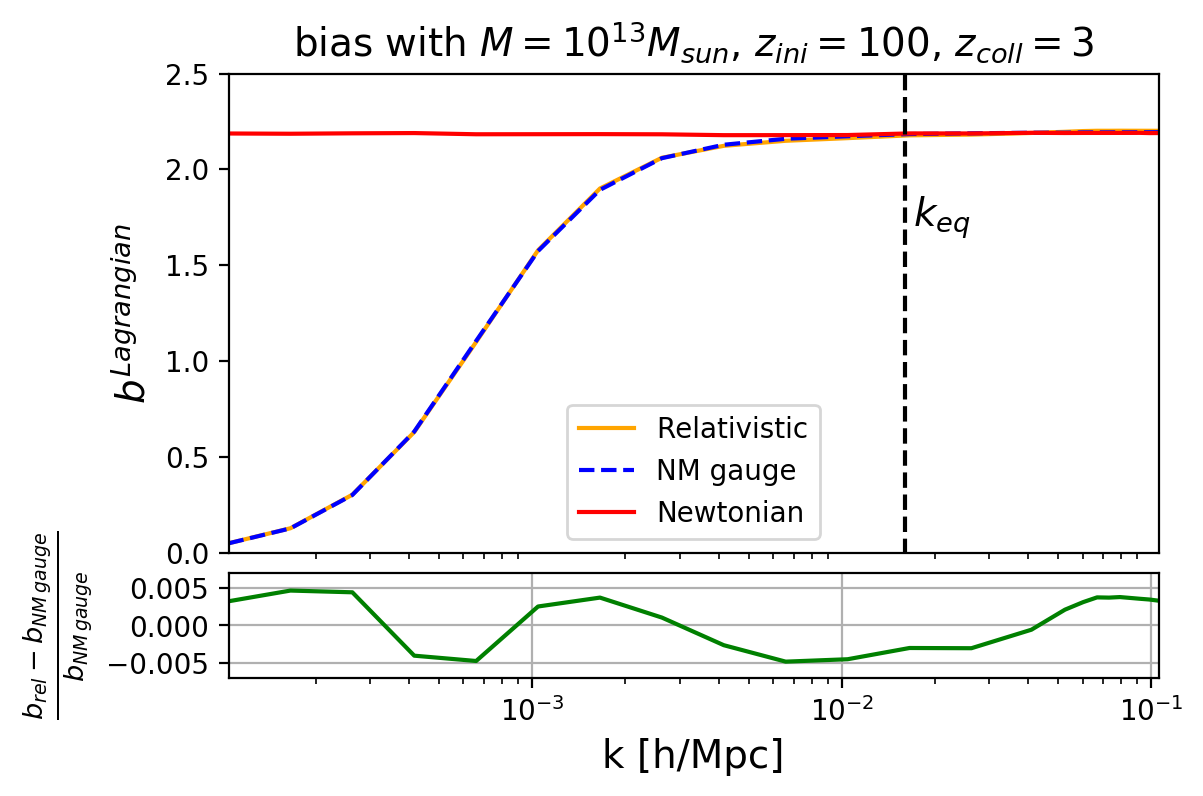}
		\caption{The top plot displays the different methods of determining the halo bias. The orange and the red lines are the results of the relativistic and Newtonian simulation, respectively. The black dashed vertical line marks the matter radiation equality scale. The blue dashed line represents the transformation from Newtonian to relativistic bias (see Equation 4.5 of \cite{Fidler:2018geb}) in the Newtonian motion gauge framework. The relative precision of both approaches is illustrated in the bottom plot.}
		\label{fig:NM-compare}
	\end{figure}

	We crosscheck our results against the semi-analytical prediction of the bias in the Newtonian motion gauge framework \cite{Fidler:2018geb}. To do this we repeat our analysis assuming a simpler Newtonian dynamics, finding as expected a scale-independent bias. This confirms that our analysis does not introduce spurious scale dependancies. We then forecast the GR-bias from the Newtonian one as described in \cite{Fidler:2017pnb} and compare the results against our full relativistic run. The results are shown in Figure \ref{fig:NM-compare} where our relativistic computation matches the prediction at the permille level. 
	Note that the Newtonian bias employs identical assumptions for the mass function and therefore this part of our analysis cannot be tested by this comparison. However, this cross-check provides an excellent confirmation for our relativistic geodesic equation approach (Eq.~\ref{GR:shell}) to the shell evolution. It should further be noted that the Newtonian motion framework does in principle not rely on spherical shell simulations, even though our Newtonian simulations do. 
	\section{The impact of neutrinos on structure formation}
	
	We finally have all the tools required to study the impact of neutrinos on small scale structure formation. On the one hand, neutrinos affect the long modes evolution in which the shell collapses. This effect is taken into account via our geodesic equation for dark matter and does not require any further modifications in our analysis. Furthermore, neutrinos may also directly contribute to the process of the gravitational collapse should they be slow enough to be captured. Note that neutrinos do not have to follow the collapse completely, it is sufficient if a lose overdense cloud of neutrinos accumulates around the forming structure.
	
	To study the neutrino capture we employ three different approximations. In the first case we assume that neutrinos do not cluster at all and do not contribute actively to the collapsing shell. We call this case 'no collapse'. The second approximation is to assume that neutrinos collapse as cold matter as soon as they become non-relativistic, neglecting any residual pressures. We define the effective density
	\be
	\rho_{coll} = \rho_\nu - 3 p_\nu\,,
	\ee
	and add the corresponding (time-dependent) contribution to the shell. While the first case provides a lower bound to the impact of neutrinos, this scenario describes the maximum impact neutrinos can have and we label it 'maximum collapse'. In our third approach we go beyond the capabilities of the closed Universe ansatz and describe a more realistic model where the complex phase-space of the neutrinos is simulated. As for the dark matter we describe the local neutrino density using a spherical symmetric fluid representing small scale perturbations in the neutrino distribution that are sourced from the gravitational potential of the dark matter shell. This case is labeled 'realistic collapse'.
	
	In general neutrinos may have a relative velocity with respect to the collapse and posses higher kinetic moments. These effects are however suppressed by the assumption that the large mode is significantly larger than the scale of collapse. We have tested that including a relative velocity between neutrinos and the collapsing shell only slightly affects the collapse dynamics. However, all of these extensions significantly complicate the problem as they explicitly break the spherical symmetry of the setup. For this reason we choose to neglect them here and work only with the neutrino density and pressure. 
	
	\begin{figure}[htbp]\label{fig:neutrino-coll}
		\centering
		\includegraphics[width=0.95\textwidth]{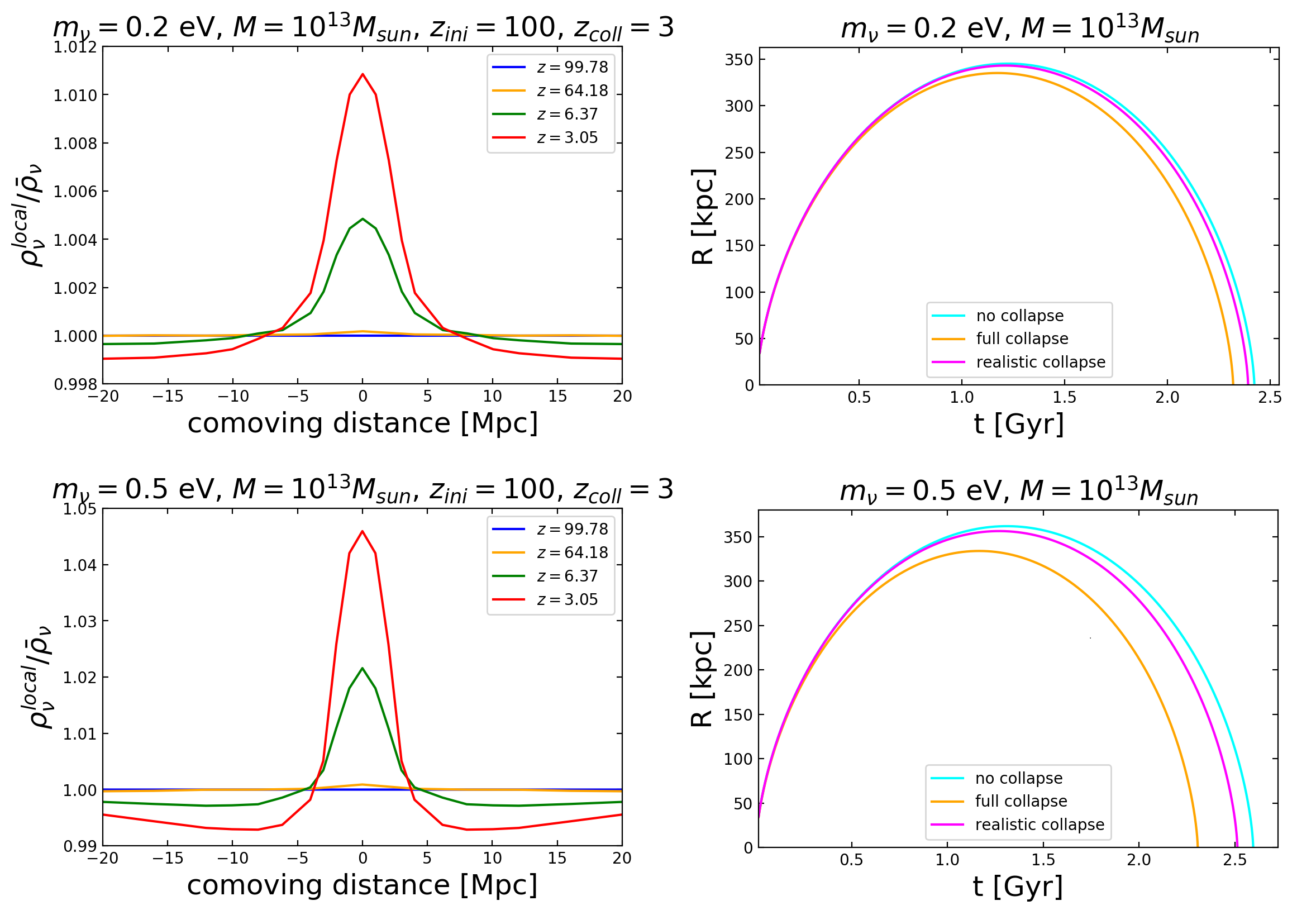}
		\caption{Impact of the different approximations for the capture of neutrinos by the small scale structures, for two neutrino mass sums, $m_\nu=0.2$ eV (top panels) and $m_\nu=0.5$ eV (bottom panels). On the right hand side we see the collapse histories for our three different cases of modeling neutrinos. On the left side we show the response of the neutrino density to the cold dark matter mass shell. Initially the neutrinos are not affected since they are still too fast to be captured. However, at later times, especially if neutrinos are more massive, they start clustering loosely around the shell.}
	\end{figure}
	We show our results in Figure \ref{fig:neutrino-coll} for two different neutrino masses. The plots on the left show the evolution of a small scale neutrino overdensity that accumulates around the collapsing shell. Initially the hot neutrinos do not respond to the overdensity since their pressure prevents the collapse, but after the non-relativistic transition they start to accumulate around the shell. Note that the overall neutrino concentration stays relatively close to the background value. For a combined mass of $0.2$ eV or smaller we find that neutrinos are only captured at the later stages of the collapse and then have a small impact on the collapse. This is not true for larger neutrino masses, which start influencing the collapse already from turnaround. The approximation of the full collapse vastly overestimates the importance of neutrinos, while the realistic model and non-collapsing neutrinos are much closer together. Still a clear difference of up to $\sim 30$\% is found depending on the neutrino mass. 
	
	Previous analyses (e.g. \cite{LoVerde:2014pxa}) assumed that the neutrino capture can be neglected for the allowed neutrino mass range, while we find that it plays an important role. This difference seems to be caused by time mismatch in the closed Universe case that incorrectly leads to the neutrino perturbations being evaluated at much earlier times. In that case we can see from Figure \ref{fig:neutrino-coll} that we would also find a negligible role of neutrinos on the collapse. 
	
	\begin{figure}[htbp]
		\centering
		\includegraphics[width=0.95\textwidth]{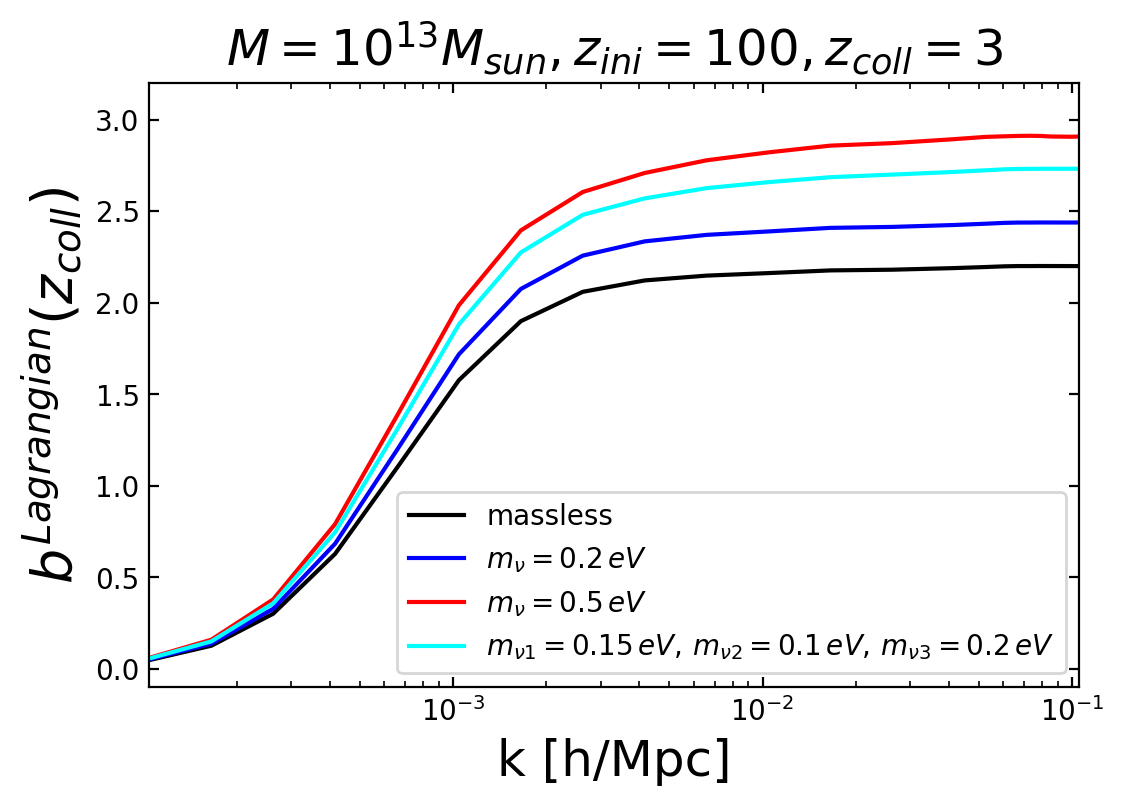}
		\caption{Neutrino bias in a massless neutrino cosmology (black line), in a massive neutrino cosmology with a single neutrino of mass $m_\nu=0.2$ eV (blue line) and $m_\nu=0.5$ eV (red line). Finally, the cyan line shows the impact of the three individual neutrino masses $m_{\nu i}$ for $i=1,2,3$.}
		\label{fig:neutrino-bias}
	\end{figure}
\begin{figure}[htbp]\label{fig:neutrino-rel}
		\centering
		\includegraphics[width=0.95\textwidth]{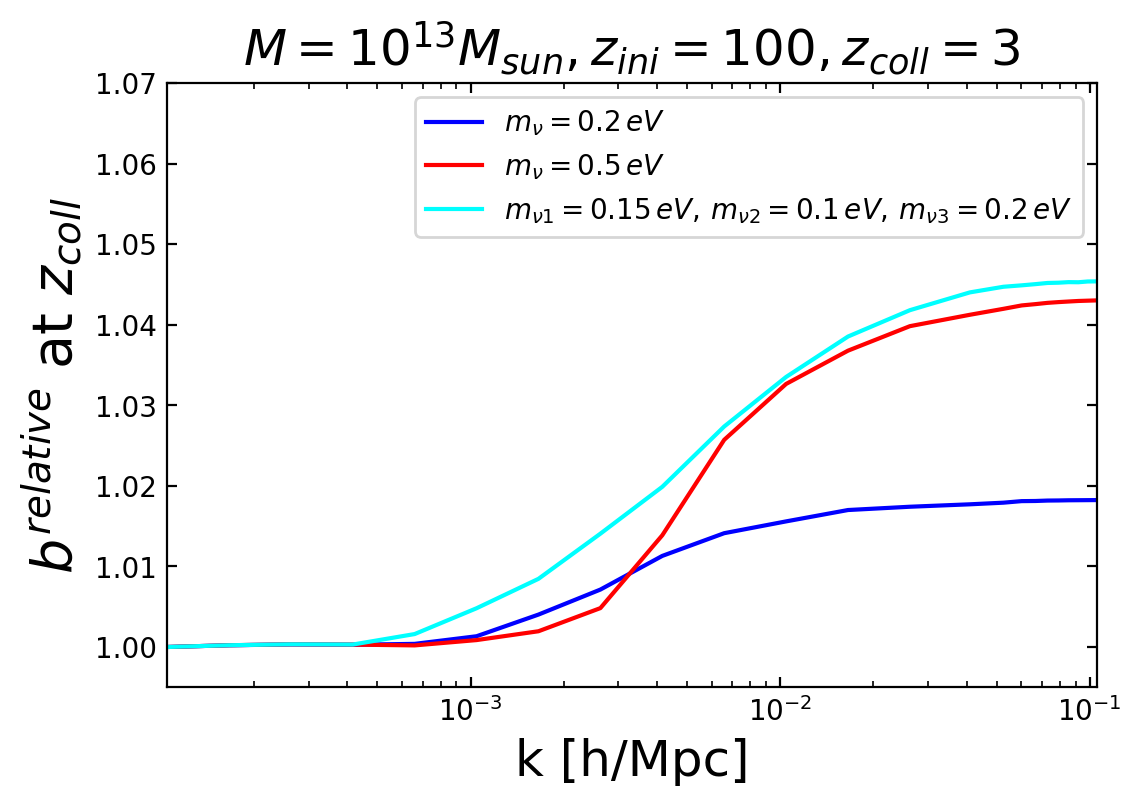}
		\caption{The bias relative to the largest scales and a neutrino-less cosmology. We see that the neutrino feature affects scales smaller than their free-streaming scale and the amplitude is mostly driven by the total neutrino mass. }
	\end{figure}

	Including the realistic neutrino capture we then repeat the analysis described in section \ref{sec:numerics} to compute the bias in cosmologies with massive neutrinos. Our results are shown in Figure \ref{fig:neutrino-bias}. We reproduce the large scale feature stemming from relativistic effects that we also found in the massless case, but depending on the neutrino mass we find a different evolution of the bias towards smaller scales. To highlight the impact of neutrinos we define a relative bias, where we compute the bias relative to a cosmology with massless neutrinos and relative to a reference scale: 
	\be
	b^{\rm relative} = \frac{b(k)}{b(k_{\rm large})} \bigg / \frac {b^{\rm m_\nu = 0}(k)}{b^{\rm m_\nu = 0}(k_{\rm large})}\,,
	\ee 
	We show this relative bias in Figure \ref{fig:neutrino-rel}. The neutrino feature develops at the free-streaming scale of the lightest neutrino while the amplitude is dominated by the total mass sum. The overall effect of neutrinos on the bias is of a few percent compared to a neutrino-less cosmology.

	In contrast to our results, in the analysis of \cite{LoVerde:2014pxa} the amplitude depends not only on the total mass, but also on the individual masses. Our results agree quantitatively with the one found in \cite{Munoz:2018ajr} where it is discussed that the difference with respect to \cite{LoVerde:2014pxa} comes from the treatment of neutrinos in the closed Universe ansatz. We agree with this interpretation and have seen that our approach of using the geodesic equation also provides an amplitude that is independent of the individual neutrino masses. 
	
	While the differences with respect to \cite{Munoz:2018ajr} are much smaller, our descriptions do not fully agree. In particular both \cite{LoVerde:2014pxa} and \cite{Munoz:2018ajr} find a scale dependent neutrino-like feature also in the massless-case that has a comparable amplitude and scale-dependance than the neutrino signature. In our analysis we find a significantly larger relativistic feature on scales beyond the matter-radiation equality scale, seen in Fig.~\ref{fig:neutrino-bias}. We believe that this difference is again rooted in the use of the closed Universe ansatz\footnote{A full comparison of the results of our work, relying on the Poisson gauge throughout, to those of \cite{LoVerde:2014pxa} and \cite{Munoz:2018ajr} is difficult, since they do not include large-scale relativistic corrections and thus do not specify the gauge.}. While \cite{Munoz:2018ajr} correctly includes the neutrino long modes in that case, a similar matching needs to be applied to the long modes of any species, except cold dark matter, i.e.~including photons. Our result agrees with the well known scale-independent bias found in numerical simulations for massless neutrino cosmologies after taking into account the respective gauge \cite{Blanton:1998aa,Sheth:1999mn,Fidler:2018geb}.
	
	%%%%%%%%%%%%%%%%%%%%%%%%%
	\section{Conclusions}\label{sec:conclusions}
	In this work we have investigated the relativistic large scale halo matter bias in
	mixed dark matter cosmologies. To do so we have developed a framework for
	calculating the collapse of matter in a relativistic setup in the Poisson gauge. Our approach takes into account consistently relativistic corrections in the weak field limit of GR that is valid from the largest super-horizon scales to the small scales on which individual halos form. Our main simplification is assuming a spherical collapse, making the problem simple enough to solve it using a semi-analytic framework. Even though this will only describe small fraction of the realistic collapses, it should not fundamentally change the impact that a long mode perturbation has on the small scale collapse, leading to an accurate bias. 
	
	We have tested our results against the predictions of the Newtonian motion gauge framework in the case of massless neutrinos and find consistent results. 
	Comparing our relativistic description for the collapse of a dark matter shell with the commonly employed closed Universe ansatz we find
	that the inclusion of Hubble friction leads to a strong discrepancy between the two approaches. This turns out to be an issue of interpretation and can be reconciled by a change of the time coordinate in a pure dark matter cosmology. In more complex cases specific matching relations between both methods have been studied in \cite{Hu:2016ssz}.
	This issue also concerns the definition of the critical density that is often based on the time evolution found in the closed Universe case. However, the spherical symmetry already causes large systematics in the computations of the mass function and fits to N-body simulations are frequently used instead. These do not require the input of a critical density and can be made compatible with both approaches.
	
	Our method allows us to incorporate neutrinos consistently, including corrections from their relativistic nature. 
	We employ a fluid description based on the Navier-Stokes equations to describe the accurate response of the neutrinos to the local shell over-density and capture their impact on the collapse without having to  rely on strong approximations. Opposed to earlier results in \cite{LoVerde:2014pxa}, we find that the neutrino accumulation on the small scales provides a significant correction to the evolution of the dark matter shell and consequently the bias, especially in more massive scenarios.
	
	We find that the relativistic bias is generally scale dependent and has a strong feature at the scale of matter-radiation equality independent of the neutrino mass. It is generated from the modified time-evolution of modes that enter the horizon already during radiation domination. This feature was already discussed in \cite{Fidler:2018geb} and is gauge dependent. If we transform our results to the synchronous gauge this feature exactly cancels providing the well-known scale-independent bias. 
	
	Including massive neutrinos we are able to resolve the characteristic
	features from their free-streaming scales. Even for small masses the bias
	shows significant differences compared to a massless neutrino cosmology. 
	The impact of the massive neutrinos depends mostly on the combined neutrino mass, while the shape of the scale-dependent feature is sensitive to the individual masses. 
	The overall relative impact of neutrinos is found to be at the few percent level.
	
	This impact of neutrinos on the bias is relevant for future surveys, where the neutrino bias partially offsets the neutrino induced suppression of the matter power spectrum. Furthermore the impact of neutrinos on the bias is sensitive to the individual masses, in principle allowing us to constrain them individually and not only gaining information on the combined mass. However, this difference remains small and it will be a challenge to observe it with sufficient precision.

	\section*{Acknowledgements}
	
	We thank Nils Sch\"oneberg for helpful discussions.

	\bibliography{references}
	\bibliographystyle{JHEP}

\end{document}